# Novel graphene/Sn and graphene/SnO$_x$ hybrid nanostructures: Induced superconductivity and band gaps revealed by scanning probe measurements

*András Pálinkás*[§,†,*], *György Molnár*[§], *Gábor Zsolt Magda*[§,†], *Chanyong Hwang*[‡,†], *Levente Tapasztó*[§,†], *Peter Samuely*[¶], *Pavol Szabó*[¶], *and Zoltán Osváth*[§,†,*]

[§]Institute of Technical Physics and Materials Science (MFA), Centre for Energy Research, Hungarian Academy of Sciences, 1525 Budapest, P.O. Box 49, Hungary

[†]Korea-Hungary Joint Laboratory for Nanosciences (KHJLN), P.O. Box 49, 1525 Budapest, Hungary

[‡]Center for Nano-metrology, Division of Industrial Metrology, Korea Research Institute of Standards and Science, Yuseong, Daejeon 305-340, Republic of Korea

[¶]Centre of Low Temperature Physics @ Institute of Experimental Physics (IEP), Slovak Academy of Sciences (SAS) and P. J. Šafárik University, 040 01 Košice, Slovakia

**Abstract**

The development of functional composite nanomaterials based on graphene and metal nanoparticles (NPs) is currently the subject of intense research interest. In this study we report the preparation of novel type of graphene/Sn and graphene/SnO$_x$ ($1 \leq x \leq 2$) hybrid nanostructures and their investigation by scanning probe methods. First, we prepare Sn NPs by evaporating 7 – 8 nm tin on highly oriented pyrolytic graphite substrates. Graphene/Sn nanostructures are obtained by transferring graphene on top of the tin NPs immediately after evaporation. We show by scanning tunnelling microscopy (STM) and spectroscopy (STS) that tin NPs reduce significantly the environmental *p*-type doping of graphene. Furthermore, we demonstrate by low-temperature STM and STS measurements that superconductivity is induced in graphene, either directly supported by Sn NPs or suspended between them. Additionally, we prepare $SnO_x$ NPs by annealing the evaporated tin at 500 ºC. STS measurements performed on hybrid graphene/$SnO_x$ nanostructures reveal the electronic band gap of $SnO_x$ NPs. The results can open new avenues for the fabrication of novel hybrid superconducting nanomaterials with designed structures and morphologies.

---
[*]Corresponding author. Tel: +36 1 392 2222 / 1157. E-mail: andras.palinkas@energia.mta.hu (András Pálinkás).
[*]Corresponding author. Tel: +36 1 392 2222 / 3616. E-mail: zoltan.osvath@energia.mta.hu (Zoltán Osváth).

1. Introduction

Graphene has attracted substantial research interest due to its outstanding electronic [1], optical [2], mechanical [3], thermal [4], and electrochemical [5] properties. Additionally, the large surface area of graphene (2630 $m^2g^{-1}$) [5] makes this two-dimensional material an attractive choice as the matrix for different nanocomposites. Hybrid nanostructures integrating graphene with metallic or semiconducting nanoparticles (NPs) could potentially display versatile and tunable properties, as well as novel or enhanced functionalities arising from the synergy between the properties of graphene [6] and those of NPs [7, 8]. The graphene/nanoparticle hybrids can have many potential applications such as in energy harvesting and storage, transparent conducting films, chemical- and biosensors, or drug delivery [9]. In particular, Sn NPs attached to graphene has been proposed as a most promising anode for lithium ion batteries [10]. Here, graphene acts as the matrix material, which is able to constrain the volume expansion of Sn upon the formation of lithium rich alloys [10]. Mixed Sn-based nanoparticles (Sn, SnO, $SnO_2$) on graphene was shown to exhibit excellent electrochemical performance in terms of specific capacity and cycling stability [11, 12, 13]. Enhanced electrochemical performance was also obtained by the synthesis of layer-by-layer (graphene/Sn/graphene) 3D nanocomposites [14, 15, 16, 17] or by using carbon coated Sn NPs embedded in graphene [18] as anode material. Graphene decorated with tin or tin-oxide NPs are promising candidates also as field emitters [19, 20] or gas sensors [21, 22]. Furthermore, it was shown by charge transport measurements that tin decorated graphene became a superconductor at low temperatures [23, 24], and the superconducting phase transition could be controlled electrostatically [23, 24]. In order to better understand the interaction between tin and graphene, a detailed microscopic investigation is needed. As far as we know, the characterization of graphene/Sn hybrids by scanning tunnelling microscopy (STM) and spectroscopy (STS) is currently lacking from the literature. In this work we synthesize graphene covered Sn nanoparticles and characterize them by atomic force microscopy (AFM), and STM/STS at room- and low temperatures. We show that metallic Sn NPs induce electrostatic doping by transferring electrons to graphene. Subkelvin STM/STS measurements give spectroscopic evidence of induced superconductivity in graphene in the proximity of Sn NPs. We also demonstrate that the graphene cover prevents the oxidation of tin NPs. Graphene/SnO$_x$ hybrid nanostructures were also prepared and the electronic band gap of SnO$_x$ NPs was measured by STS.

2. **Experimental**

Tin was evaporated (7 – 8 nm) onto freshly cleaved highly oriented pyrolytic graphite (HOPG) surface, at a rate of 0.1 nm/s and background pressure of $5\times10^{-7}$ mbar. Due to the low wetting properties, the evaporated tin film self-organized into nanoparticles with heights of 15-30 nm and lateral dimensions of 50-100 nanometres. The NPs were covered with large-area graphene grown by chemical vapour deposition on a mechanically and electro-polished copper foil, as described elsewhere [25]. The transfer of graphene was performed using thermal release tape, and a copper etchant mixture consisting of CuCl$_2$ aqueous solution (20%) and hydrochloric acid (37%) in 4:1 vol ratio was used to remove the copper foil. After the etching procedure, the tape holding the graphene was rinsed in distilled water, then dried and pressed onto the tin covered HOPG surface immediately after opening the evaporation vacuum chamber. The tape/graphene/tin/HOPG sample stack was placed on a hot plate and heated to 100 ºC, above the release temperature (90 ºC) of the tape. Thus the tape was easily removed, leaving behind the graphene on top of the tin NPs (and on the HOPG). The graphene-covered tin NPs were investigated by STM and STS, using a DI Nanoscope E operating under ambient conditions. We used mechanically-cut Pt/Ir (90/10%) tips for the room temperature STM experiments. Each STS spectrum is an average of 10 measurements at the same tip position, to avoid random features. The sample was annealed in electric furnace at 500 ºC for 30 min in Ar atmosphere. AFM measurements were performed in both Tapping and PeakForce® mode using a MultiMode 8 from Bruker. The low temperature STM/STS experiments have been performed at T = 500 mK in a $^3$He STM system of IEP SAS in Kosice. The STM tips for the low temperature measurements were formed in-situ by controlled collision of a gold tip to a clean Au surface. This method enables preparing of atomically sharp tips at cryogenic temperatures [26].

3. **Results and discussion**

The Sn evaporated on HOPG surface organizes into non-percolating nanoparticles, as a result of poor wettability of graphitic surfaces by tin [24, 27]. PeakForce AFM measurements of such tin NPs partially covered with graphene are shown in Figure 1 (sample "*A*"). In addition to topographic data (Fig. 1a), adhesion forces from the same area are also collected (Fig. 1b). The blue lines drawn manually on both topography and adhesion maps demarcate graphene covered areas from non-covered regions. Generally, lower adhesion to the AFM tip is measured on non-covered Sn NPs, as shown by the area with predominantly dark contrast in Fig. 1b. Here, small areas with light contrast are intercalated, which correspond to the HOPG substrate without NPs. Such small areas without NPs occur frequently in both covered and non-covered regions (Fig. 1a, darker contrast).

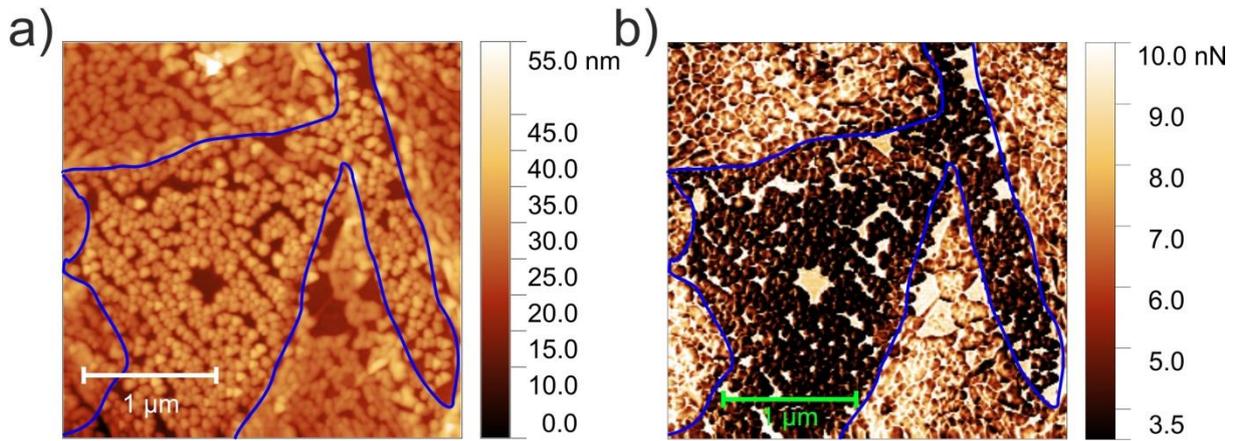

*Figure 1. PeakForce AFM topography and adhesion maps of graphene/Sn NPs. (a) Evaporated tin film (7 nm) on HOPG substrate, self-organized into nanoparticles with heights of 15-30 nm and diameters of 50-100 nm. The area is partially covered with graphene. (b) Adhesion map of the same area. Graphene covered parts show higher adhesion to the AFM tip (lighter contrast) than non-covered parts (darker contrast). The edges of graphene are marked with blue solid line on both images.*

We investigated by STM and STS the effect of Sn nanoparticles on the local density of states (DOS) of graphene. An area completely covered with graphene is shown in Figure 2a. Here, closely spaced Sn NPs can be observed. The presence of graphene is demonstrated by the atomic resolution image obtained on a covered tin nanoparticle (inset and red symbol in Fig. 2a). Note that the graphene is bridging the closely spaced, neighbouring NPs, and these graphene parts are suspended. STS measurements were carried out on both suspended and Sn supported graphene, as well as on the graphene covered HOPG (green symbol in Fig. 2a). The corresponding *dI/dV* curves are shown in Figure 2b. The STS spectra show typical V-shaped DOS characteristics [28, 29] for the graphene covered parts. The Dirac point of graphene on HOPG is measured at 130 mV above the Fermi energy ($E_F$), showing significant *p*-doping (Fig.

2b, green dash-dotted line). This is attributed to environmental doping which could originate from airborne dopants (e.g. H$_2$O, OH) adsorbed on graphene [30, 31]. In contrast, for Sn supported graphene the *dI/dV* minimum is close to $E_F$ (Fig. 2b, red solid line), i.e. considerably shifted compared to graphene on HOPG. Estimating the DOS near $E_F$ with $N(E_F) = 2|E_F|/(\pi\hbar^2 v_F^2)$ [32, 33], the observed shift implies an electron transfer of $\Delta n = 1.23 \times 10^{12} cm^{-2}$ from the Sn NP to graphene (assuming $v_F = 10^6 m/s$), which reduces significantly the pre-existing environmental *p*-doping. As for the suspended graphene, STS spectra measured on two different positions (Fig. 2a, black and blue symbols) are shown in Figure 2b (black dashed and blue dotted lines, respectively).

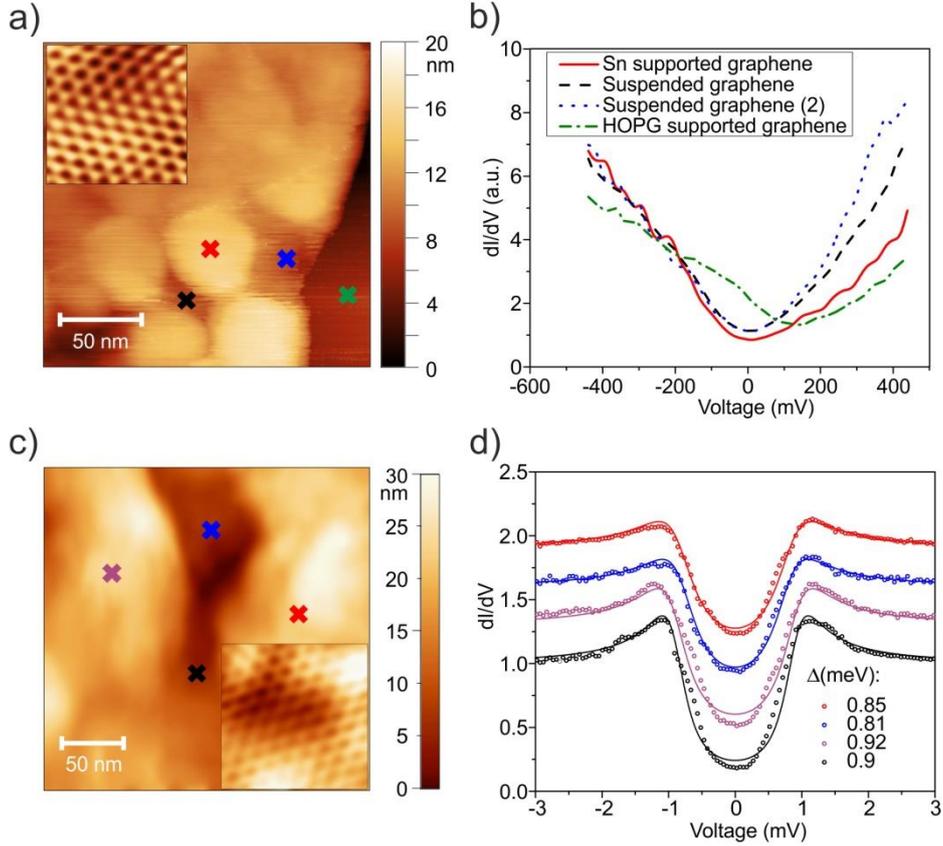

*Figure 2. STM and STS investigation of graphene/Sn NPs. (a) and (c) STM topography of graphene covered areas measured at T = 300 K and T = 0.5 K, respectively. Individual Sn NPs below the graphene can be easily observed (U= 1 V, I= 300 pA). Atomic resolution images of Sn supported graphene are shown in the insets. (b) and (d) STS spectra measured at T = 300 K and T = 0.5 K, respectively. The spectra were measured at the positions marked with coloured symbols in a) and c), as follows: Sn supported graphene (red, purple), suspended graphene (black, blue), graphene on HOPG (green). The spectra in (d) have been normalized for their values at V = 5 meV and shifted along the vertical axis for clarity. The symbols plot the experimental data, the solid lines are the fitting curves for the Dynes model.*

Interestingly, the Dirac point of suspended graphene is obtained at the same bias voltage as for Sn supported graphene, which shows that the electrostatic doping of Sn supported graphene prevails in the suspended parts as well.

A similar sample (sample "B", see Figure S1, Supplementary information) was investigated by means of low-temperature STM at $T$ = 0.5 K when tin is in the superconducting state ($T_c$ ~ 3.7 K). The area shown in Fig. 2c with two Sn NPs is fully covered by graphene. The inset presents a zoom to the graphene atomic lattice. Tunnelling spectra have been taken at the four indicated positions. The red and purple crosses mark two positions on top of the Sn NPs while the blue and black ones indicate suspended graphene in between the Sn NPs. The *dI/dV* spectra depicted in Fig. 2d with symbols clearly show superconducting density of states with a superconducting gap at $T$ = 0.5 K, evidencing that superconductivity has been induced in both Sn supported and suspended graphene. Spectra taken in many different points (not shown here) always display a gap. However, the measured spectra also display deviations from the standard Bardeen-Cooper-Schrieffer (BCS) DOS. The superconducting DOS is significantly broadened with in-gap states on the order of 30% of the *dI/dV* at higher bias voltage. The coherence peaks are also heavily suppressed. Notably, they are very similar to the spectra taken on epitaxial multilayer graphene in proximity to a superconducting aluminum thin film at distances around hundred nanometres [34]. Such broadening of the spectra is a signature of strong pair breaking effect [35, 36]. Formally, the spectra can be very well fitted by the Dynes modification [37] of the BCS DOS $N_S(E) = Re\left(\frac{E}{\sqrt{E^2-\Delta^2}}\right)$, with the complex energy $E = E - i\Gamma$, where $\Gamma$ is the spectral broadening parameter and $\Delta$ is the superconducting energy gap [37]. The fitting curves are shown in Fig. 2d with solid lines, the corresponding $\Delta$'s are in the legend. The gap values obtained from fitting are randomly scattered between 0.6 and 0.9 meV among various places on graphene while the broadening parameter $\Gamma$ was around 0.3-0.4$\Delta$. Our results indicate that the superconducting energy gap induced from Sn NPs into graphene by proximity effect is even enhanced in comparison with the Sn bulk value, reaching between 100 and 150 % of $\Delta_{bulk}$ = 0.57 meV. Recently, a systematic study of the gap value as a function of the size of isolated Sn NPs was performed by Bose *et al.* [38]. For Sn NPs of heights between 5 and 30 nm they obtained a distribution of gap values which is very similar to our observations, including also the significant broadening of the spectra described by the $\Gamma$ parameter. The large enhancement and fluctuating size of the superconducting energy gap was theoretically explained by the quantum size effect in zero-dimensional superconductors. In a small particle the discretization of energy levels arises from quantum confinement. In superconductivity only the levels within the pairing region around the Fermi energy have significance. The number of levels within this region fluctuates depending on the position of the Fermi level, which causes the fluctuation of the gap size. Since the sizes of Sn nanoparticles in the Bose experiment reasonably match that of our Sn NPs (see also Fig. S1), we ascribe our observed gap distribution to the above mentioned quantum size effect. Moreover, the gap fluctuations result in pair breaking, which naturally explains the presence of a broadening parameter $\Gamma$.

Next, we annealed the graphene/Sn sample "*A*" at 500 ºC, in order to improve the adhesion of graphene to nanoparticles. Tapping AFM images measured after annealing are shown in Figure 3. Here, a significant difference between graphene covered and non-covered regions (demarcated with dashed line in Fig. 3a) can be observed. Interestingly, the graphene covered Sn NPs lost their initial shape and transformed into various new structures, even nanorods (Fig. 3b, blue arrows). The NPs merged together to form larger objects which sometimes have surprisingly well-defined edges enclosing 120º (Fig. 3b, blue lines). In contrast, the structure of non-covered NPs did not change, as observed in the AFM image of Figure 3a (bottom right part). This finding is attributed to the presence (absence) of graphene, which acts as a protective layer. Covered tin NPs remain metallic and they melt during annealing. The melting point of tin is at around 230º, however this value is even lower for nanoparticles [39].

While the tin melts, the graphene cover shrinks, due to its negative thermal coefficient [40], which drives the tin to aggregate into various nanostructures. The straight edges enclosing 120º are determined by the crystalline orientations (zig-zag, armchair) of graphene (see also Fig. 4).

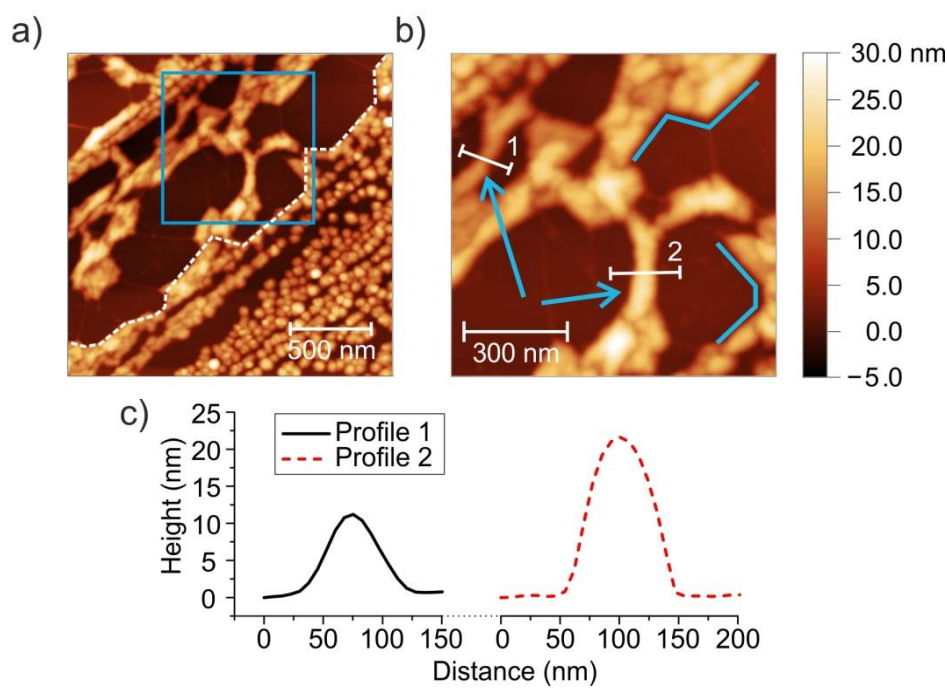

*Figure 3. AFM topography images after annealing at 500 ºC. (a) Graphene covered areas and non-covered tin nanoparticles (bottom right) are demarcated with dashed line. Non-covered tin nanoparticles preserved their sphere-like shapes, while the structure of tin under graphene transformed. (b) Higher resolution AFM image of the graphene covered squared area in (a). Tin nanorods formed under the graphene (marked with blue arrows). Larger nanostructures also formed with well-defined edges enclosing 120º (blue lines). (c) Height profiles of the nanorods measured at the line sections "1" and "2" in (b).*

Note that the annealing was performed three month after the initial tin evaporation, and graphene was able to protect efficiently the Sn NPs from oxidation in this time frame (the sample was stored under ambient conditions). On the other hand, non-covered NPs exposed to air slowly oxidized, and thus they did not melt during annealing at 500 ºC (the melting points of SnO and $SnO_2$ are 1080 ºC and 1630 ºC, respectively). The sphere-like shape of NPs was preserved.

STM investigations performed on the graphene covered annealed sample show a network of less structured Sn nanoparticles (Figure 4a), in accordance with the AFM data.

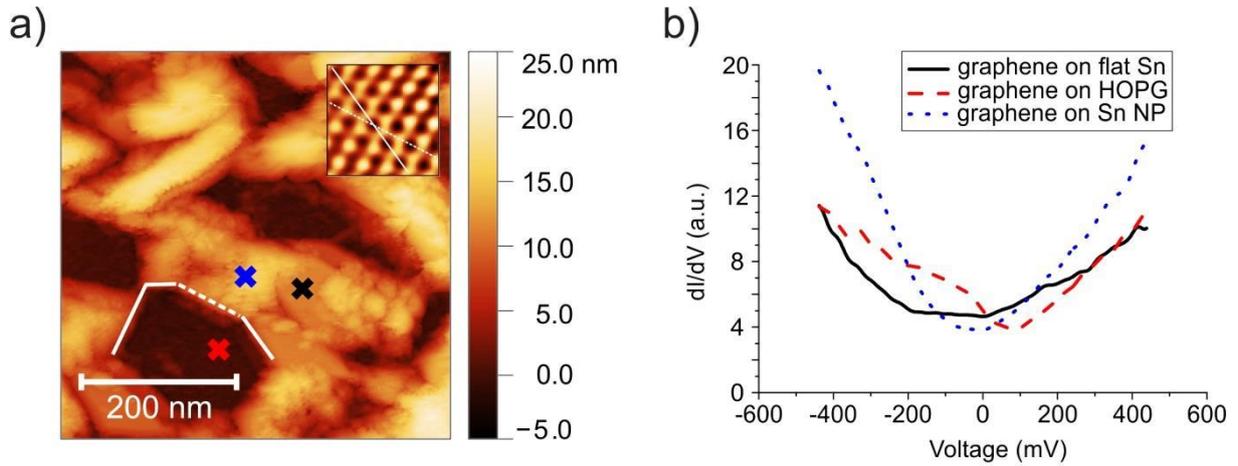

*Figure 4. STM topography (a) and STS spectra (b) measured after annealing at 500 ºC. (a) The STM image shows an area completely covered with graphene. The edges of Sn nanostructures are sometimes determined by the zig-zag (white line) and armchair (dashed white line) orientations of graphene, as shown by the atomic resolution image in the inset, measured on HOPG supported graphene (red symbol). The spectra in (b) correspond to: graphene on HOPG (red dashed line), graphene on Sn NP (blue dotted line), and graphene on flat Sn plateau (black solid line). These spectra were measured in three different positions marked in (a) with red, blue, and black symbols, respectively.*

STS measurements reveal doping characteristics similar to the as-prepared graphene/Sn (Fig. 2c), as shown in Figure 4b. Here, the Dirac point of graphene on HOPG is measured at 80 mV (Fig. 4b, red dashed line), while for Sn supported graphene the tunnelling conductance minimum is obtained at -35 mV (Fig. 4b, blue dotted line), showing slight *n*-type doping. Additionally, we observed several flat Sn plateaus (Fig. 4a, black symbol) imbedded in the nanoparticle network. *dI/dV* spectra obtained on these Sn plateaus show metallic (flat) DOS at negative bias voltages (near $E_F$). This indicates strong Sn-graphene interaction, where graphene loses its V-shaped DOS characteristics. This is similar to what we observed recently on graphene/Au(110) [25], and it is attributed to the hybridization of carbon $\pi$-states with the metallic substrate [41]. Additionally, some of the Sn nanostructures have well defined straight edges, as shown already in the AFM images (Fig. 3). Here, in Figure 4a, we observe edges that enclose either 120º or 150º. Using atomic resolution STM (inset of Fig. 4a) we demonstrate that these edges are aligned along the graphene zig-zag (Fig. 4a, white solid lines) or armchair (Fig. 4a, white dashed lines) orientations.

For comparison, we prepared a graphene/SnO$_x$ nanoparticle hybrid material as follows: after evaporating 6.5 nm of Sn on HOPG substrate the sample was taken out from the evaporator and introduced into an electric furnace, where it was annealed at 500 °C for 30 min in Ar atmosphere. Graphene was transferred onto the NPs only six weeks after annealing. During this time frame the sample was left under ambient conditions. A second annealing at 500 °C was performed after the graphene transfer. AFM measurements of the annealed sample show both graphene covered and non-covered nanoparticles (Figure 5a).

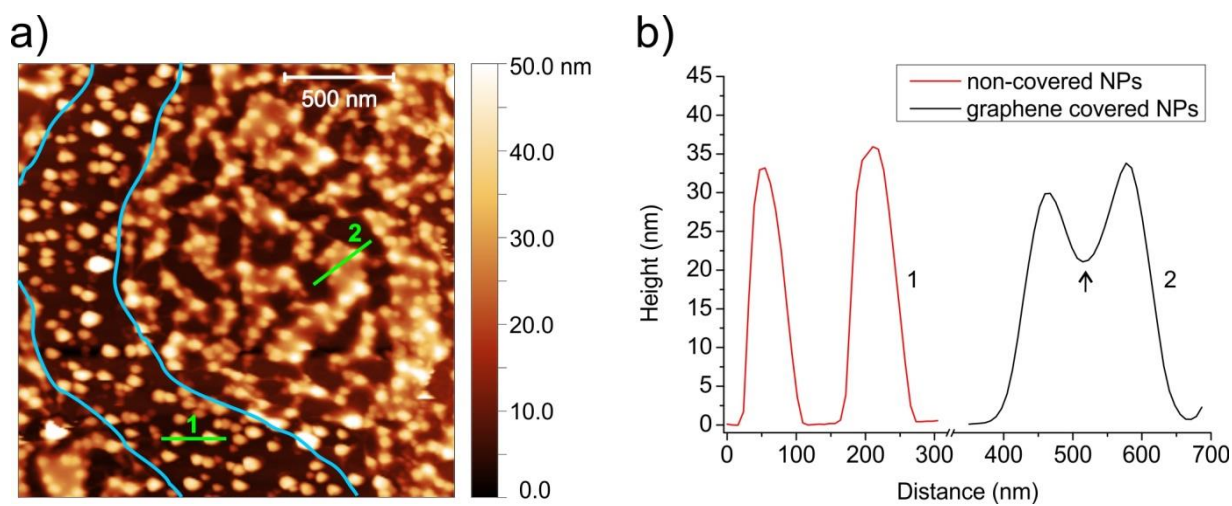

*Figure 5. PeakForce AFM measurements of graphene/SnO$_x$ nanoparticles. (a) Topographic image showing both graphene covered and non-covered NPs. Non-covered NPs are in the area between the two blue lines. The line sections 1 and 2 (green lines) are taken on two non-covered and two graphene covered NPs, respectively. (b) Height profiles corresponding to the line sections 1 and 2 in a). Graphene suspended between nanoparticles is denoted by an arrow in the height profile 2.*

In contrast to the results shown in Fig. 3, here the size and shape of covered and non-covered NPs is similar. This shows that the graphene covered NPs did not melt during annealing, so they are oxidized. Many nanoparticle groups are observed where graphene is suspended between NPs. This is shown for example in Figure 5b, where height profiles taken on non-covered and graphene covered NPs are compared.

Finally, we investigated the graphene/SnO$_x$ hybrids by STM and STS. The non-covered oxidized nanoparticles could be imaged only by using large bias voltages of 2.5 V. Such high voltages were needed in order to overcome the electronic band gap of SnO$_x$. When using lower bias voltages (U = 1.5 V, see Figure S2, Supplementary information), the STM tip got into mechanical contact with the NPs due to insufficient conductance, and pushed the NPs out of the scan window. In contrast, graphene covered SnO$_x$ NPs could be well investigated using lower bias voltages also, since graphene immobilizes the NPs and good tunnelling conductance is achieved through it. We performed STS measurements on individual graphene covered SnO$_x$ nanoparticles. The *I-V* characteristics at lower bias voltages (-1 V < U < 1 V) are very similar to the ones measured on graphene/HOPG (Figure 6).

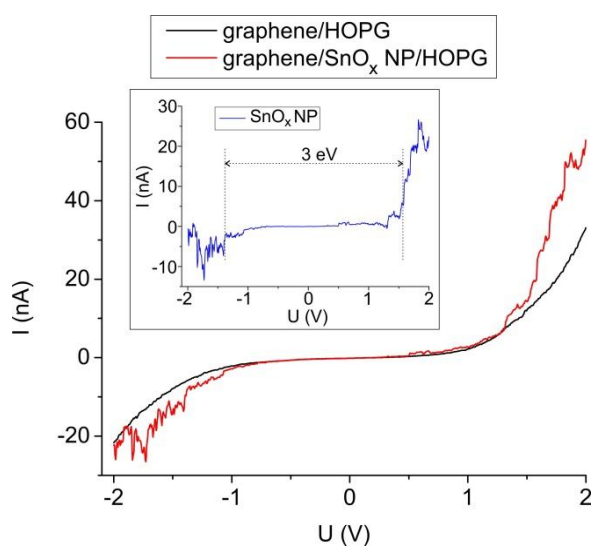

*Figure 6. STS spectra of graphene/SnO$_x$ hybrids*. *Tunneling spectra measured on graphene/HOPG (black line) and a graphene covered SnO$_x$ nanoparticle (red line). Subtracting the graphitic contribution we obtain the tunnelling conductance through the nanoparticle (inset, blue line), which shows a band gap of 3 eV.*

This shows that in this voltage interval the tunnelling current is carried by the graphene overlayer into the HOPG substrate, and there is no conductance through the SnO$_x$ NPs. The current increases and starts to differ considerably from the graphitic behaviour only at higher bias voltages, typically U > 1.5 V, where the bottom of the conduction band is reached, and charge transport through the SnO$_x$ NPs emerges, as a parallel channel for current flow. Similarly, the tunnelling conductance increases also at U < –1.4 V, i.e. in the valence band. The STS measurements revealed band gaps of 2.7 – 3.2 eV, which are better observed by subtracting the graphitic contribution of the current (inset of Fig. 6). These values are lower than the band gap of SnO$_2$ (3.6 eV [42]). Nevertheless, it has been shown recently that the band gap of SnO$_2$ NPs can be as low as 3.27 eV [43] for particle size ranging between 10 and 30 nm. The electronic band gap of Sn$_3$O$_4$ is in the range of 2.4 – 2.76 eV [44, 45]. Taking into account that calculations show a gap of 2.98 eV for Sn$_2$O$_3$ [46], and similar values were obtained for SnO as well [47], we think that the NPs shown in Fig. 5 can be in one or more of these oxidized phases (SnO, Sn$_3$O$_4$, Sn$_2$O$_3$, or SnO$_2$).

## 4. Conclusions

In conclusion, graphene covered Sn and SnO$_x$ (1 ≤ x ≤ 2) nanoparticles were synthesized and characterized by scanning probe methods (AFM, STM, STS). We have shown that charge transfer occurs from Sn NPs, inducing uniform doping of graphene (supported and suspended), which actually reduces significantly its environmental *p*-doping. Superconductivity induced to graphene by proximity effect from Sn NPs has been observed at low temperatures of 0.5 K. We have shown that superconductivity is induced not only in supported graphene, but also in areas where graphene is suspended. The superconducting energy gap is enhanced in comparison to bulk tin as a consequence of the quantum size effect in Sn nanoparticles. The graphene overlayer protects Sn NPs from oxidation. They melt during annealing at moderate temperatures and merge to form various larger nanostructures, including nanorods. Furthermore, we have shown that graphene facilitates the STM investigation of semiconducting SnO$_x$ NPs by fixing them to the HOPG substrate. We measured electronic band gaps of 2.7 – 3.2 eV for the $SnO_x$ nanoparticles investigated by STM and STS. The presented facile synthesis of graphene/Sn NPs and graphene/$SnO_x$ NPs can open new possibilities for the elaboration of hybrid superconducting nanostructures and for the development of functional composite nanomaterials with designed structures and properties.

## Acknowledgment

The research leading to these results has received funding from the People Programme (Marie Curie Actions) of the European Union's Seventh Framework Programme under REA grant agreement n° 334377, and from the Korea-Hungary Joint Laboratory for Nanosciences. The NKFIH OTKA grant K-119532, the NKFIH project TÉT_12_SK-1-2013-0018, as well as the APVV project SK-HU-2013-0039, are acknowledged. C. Hwang acknowledges funding from the Nano-Material Technology Development Program (2012M3A7B4049888) through the National Research Foundation of Korea (NRF) funded by the Ministry of Science, ICT and Future Planning. P.S. and P. Sz. have been supported by the Slovak projects APVV-16-0372 and VEGA 1/0409/15. G.Z.M. and L.T. acknowledge support from the Lendület programme of the Hungarian Academy of Sciences.